\documentclass[aps, superscriptaddress, showpacs,prl, twocolumn]{revtex4-1}
\usepackage{ORI_Group_style}

\hypersetup{colorlinks,linkcolor=blue,citecolor=blue,urlcolor=blue}

\newcommand{\wmodesz}{f_z(k,z)}

\newcommand{\ak}{\aop_k}

\newcommand{\dispersion}{\omega(k)}
\newcommand{\amplitude}{\alpha_k}

\newcommand{\couplingh}{g(d,t)}

\newcommand{\vphi}{\varphi(d,t)}
\newcommand{\detuning}{\Delta(d,t)}

\newcommand{\driving}{D(t)}
\newcommand{\drivingc}{D^*(t)}

\begin{document}
\title{Remote Individual Addressing of Quantum Emitters with Chirped Pulses}
\author{S.~Casulleras}
\author{C.~Gonzalez-Ballestero}
\author{P.~Maurer}
\affiliation{Institute for Quantum Optics and Quantum Information of the Austrian Academy of Sciences, 6020 Innsbruck, Austria }
\affiliation{Institute for Theoretical Physics, University of Innsbruck, 6020 Innsbruck, Austria.}
\author{J.~J.~Garc\'{i}a-Ripoll}
\affiliation{Instituto de F\'{i}sica Fundamental IFF-CSIC, Calle Serrano 113b 28006 Madrid, Spain.}
\author{O.~Romero-Isart}
\affiliation{Institute for Quantum Optics and Quantum Information of the Austrian Academy of Sciences, 6020 Innsbruck, Austria }
\affiliation{Institute for Theoretical Physics, University of Innsbruck, 6020 Innsbruck, Austria.}

\begin{abstract}

We propose to use chirped  pulses propagating near a bandgap to remotely address quantum emitters. We introduce
a particular family of chirped pulses that dynamically self-compress to sub-wavelength spot sizes during their evolution in a medium with a quadratic dispersion relation.  We analytically describe how the compression distance and width of the pulse can be tuned through its initial parameters. 
We show that the interaction of such pulses with a quantum emitter is highly sensitive to its position due to effective Landau-Zener processes induced by the pulse chirping.
Our results propose pulse engineering as a powerful control and probing tool in the field of quantum emitters coupled to structured reservoirs.
\end{abstract}

\maketitle

An exciting platform in quantum optics, both in the microwave~\cite{Albrecht_2019,PhysRevX.9.011021,ferreira2020collapse,s41567-019-0703-5,houck-rew,s41535-020-0220-x,Winkel_2020,kim2020quantum}, and the optical 
\cite{nature02772,PhysRevLett.101.113903,RevModPhys.87.347,nphoton.2015.57, nphoton.2015.54,PhysRevLett.115.163603,Paulisch_2016,Hood10507,nature21037,PhysRevA.96.023831,Belloeaaw0297} regime, is obtained by coupling quantum emitters to photonic structures where bandgaps and dispersion relations can be engineered. In essence, these systems allow enhancing and tailoring sub-wavelength light-matter interaction and bath-mediated coupling between quantum emitters. There are multiple applications in the context of quantum simulation \cite{nphoton.2015.57, nphoton.2015.54} and computation \cite{s41567-019-0703-5,Winkel_2020,Paulisch_2016} as well as in exploring unconventional quantum optics \cite{Albrecht_2019,PhysRevX.9.011021,ferreira2020collapse,houck-rew,s41535-020-0220-x,kim2020quantum,nature02772,PhysRevLett.101.113903,RevModPhys.87.347,PhysRevLett.115.163603,Hood10507,nature21037,PhysRevA.96.023831,Belloeaaw0297}.
Most of these setups rely on, or would benefit from, the possibility of electromagnetically addressing individual quantum emitters. 
However, such addressing can be challenging due to, for instance, insufficient (\eg~sub-wavelength) separation between contiguous emitters or to phase mismatch between outside radiation and the electromagnetic modes of the structure. Even in platforms where local probes are available, such as superconducting circuits, these probes might introduce unwanted decoherence and lack the flexibility that a fully tuneable local probe could provide.
A potential route towards such individual addressing could be paved by \emph{active} electromagnetic engineering, where not only the dispersion relation but also the time-dependent state of the electromagnetic environment is tailored. 

\begin{figure}[t]
\noindent \begin{centering}
\includegraphics[scale=1]{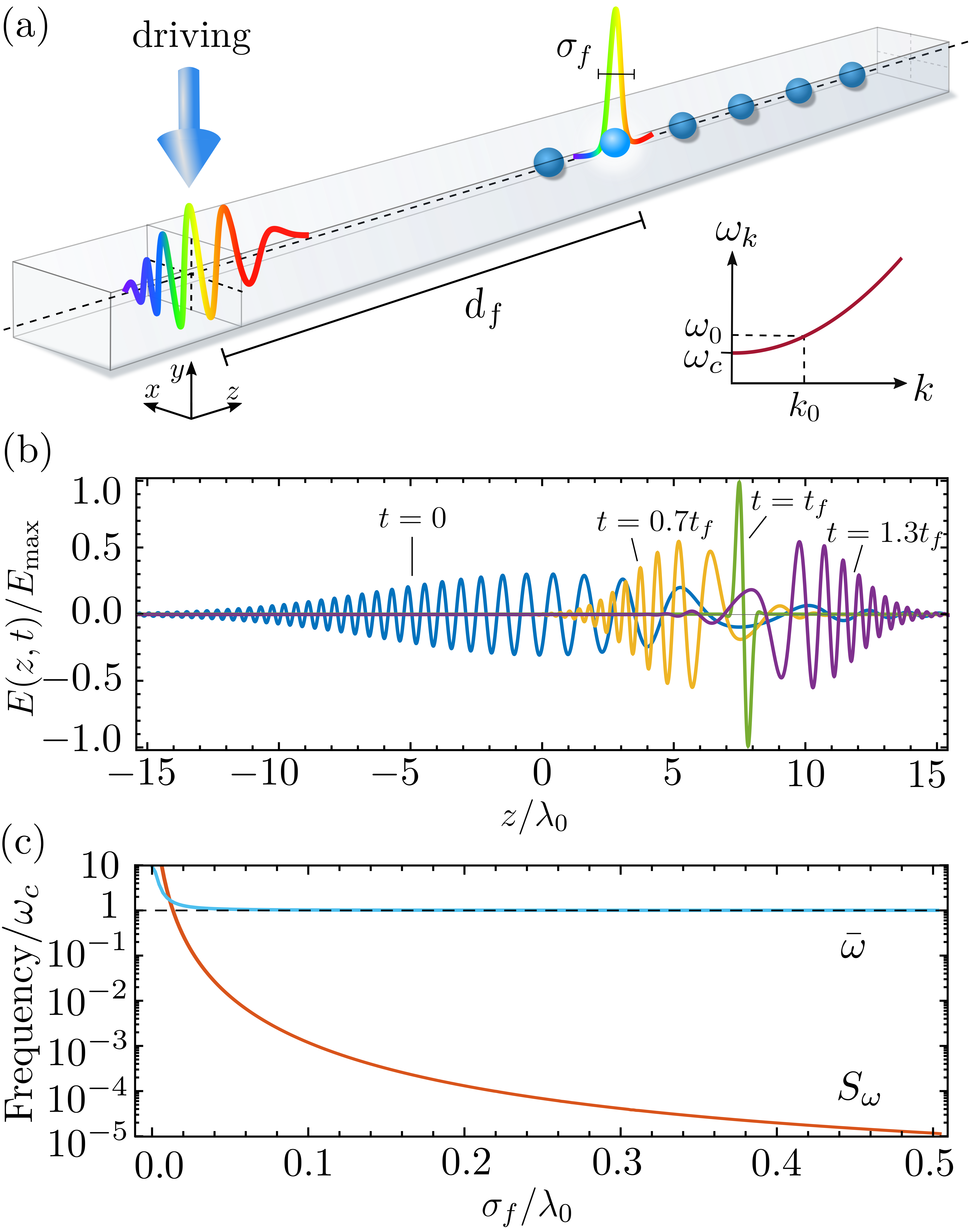}
\par\end{centering}
\caption{(a) Quantum emitters embedded in an electromagnetic waveguide. A time-dependent driving applied at the origin of the coordinate system creates a chirped self-compressing electromagnetic pulse.  At a time $t_f$ the pulse becomes compressed at a distance $d_f$ from the origin, reaching a minimum width $\sigma_f$. Inset: Quadratic dispersion relation of the waveguide.  The distribution of the pulse wavenumber along $z$ is centered around $k_0=2 \pi/\lambda_0$. (b)  Spatial profile of the electric field of the chirped pulse at different times. The electric field is normalized by its maximum value $E_{\rm max} =\max_{d,t} E(d,t)$.  (c) Mean frequency $\bar \omega$ and standard deviation $S_\omega$ (defined in the text) of the electric field pulse as a function of the compression width $\sigma_f$. Parameters used: $\omega_0/\omega_c=1.005$, $d_f/\lambda_0=7.5$, $\sigma_f/\lambda_0=0.21$, $\phi=0$. \label{fig:1} }
\end{figure}

In this paper we explore the possibility of exploiting active engineering in structured electromagnetic reservoirs. In particular, we introduce a specific family of \emph{chirped} electromagnetic pulses and show that, in a medium displaying a quadratic dispersion relation above a bandgap, their free evolution causes them to dynamically self-compress into a single, potentially sub-wavelength, spot.
Self-compression of chirped pulses using materials with nonlinear electromagnetic response (\eg~with intrinsic Kerr non-linearities) have been exploited before \cite{Silberbergs,Chernev:92,PhysRevA.49.4085}. In contrast, here we use non-linear dispersion relations that can be engineered with linear lossless materials (\eg~photonic crystals).
We then study the interaction between these chirped pulses and a quantum emitter, demonstrating the strong impact of the pulse self-compression on the dynamics of the emitter. Specifically, a quantum emitter situated at the compression spot is shown to display radically different dynamics than a quantum emitter situated at any other position. Our results therefore suggest that chirped pulses in structured electromagnetic media can be used to remotely address individual quantum emitters within an array with sub-wavelength separation (see \figref{fig:1}(a) for a schematic representation). While we discuss our results in the context of structured photonic reservoirs, our results can be extended to other implementations where bosonic excitations propagating near a bandgap couple to quantum emitters (\eg~phononic networks coupled to color centers in diamond~\cite{PhysRevLett.120.213603}).

More specifically, we consider an electromagnetic medium extended along the $z$-axis displaying an energy bandgap for electromagnetic modes propagating along $z$ with wavevector $\mathbf{k}=k\mathbf{e}_z$. The bandgap is characterized by a cutoff frequency $\omega_c$, below which there are no $z$-propagating modes. We consider that for frequencies $\omega \gtrsim \omega_c$ the dispersion relation of the propagating modes can be written as
\be
\dispersion=\omega_c+\frac{v^2}{2\omega_c}k^2.\label{dispersion}
\ee 
Here $v$ is a dimensional parameter characterizing the band curvature. 
We assume the $z$-propagating electromagnetic modes to be tightly confined in the transverse $(x,y)$ plane in order to increase the interaction with quantum emitters, as discussed later. The propagating electromagnetic modes for a given polarization can then be described by a single mode index, namely their longitudinal wavenumber $k$, and the single band \eqnref{dispersion}. 
As mentioned before, the considered electromagnetic medium can be implemented either in the microwave regime or in the optical regime.

In the medium defined above, we focus on the time dynamics of a single component of the electric field as a function of $z$ evaluated at a given position in the transverse plane, say $(x_0,y_0)$. We label such scalar electric field as $E(z,t)=2\text{Re}\{E^+(z,t)\}$. As discussed later, $E(z,t)$ is relevant to describe the electric-dipole interaction with a quantum emitter placed at $(x_0,y_0,z)$. 
The first main result of this paper is to propose and parameterize a particular family of chirped electromagnetic pulses that dynamically self-compress due to the dispersion relation given by \eqnref{dispersion}. These pulses depend on five real parameters $(k_0, d_f, \sigma_f, \phi, N)$, defined below, and can be written as
\be
E^+(z,t) \equiv |E^+(z,t)|e^{i\theta(z,t)}e^{i\phi}e^{i( k_0 z - \w_0 t)}\label{Eplus}.
\ee
Here $k_0$ is the carrier wavenumber with corresponding frequency $\w_0 \equiv \w(k_0)$ and $\phi$ is a constant phase. The amplitude of the pulse is given by
\be
|E^+(z,t)| \equiv \frac{N}{k_c \sigma(t)}
\exp \spare{-\frac{\sigma_f^2 }{2 \sigma^4(t)}\left(z-\frac{vt}{\eta}\right)^2}, \label{EplusA}
\ee
where $k_c \equiv \w_c/v$ and $\eta\equiv k_c/k_0$.
The time-dependent pulse width is given by
\be
\sigma(t)\equiv  \sqrt[4]{\sigma_f^4+\frac{s^2(d_f,t)}{k_c^4}} \label{sigma},
\ee
where $s(z,t)\equiv\eta k_c z -\w_c t$ is a spatio-temporal dimensionless function,
$\sigma_f$ is the spot size and $d_f$ the focal distance. The constant $N$ is a pulse amplitude parameter. 
The time-dependent phase in \eqnref{Eplus}, which is responsible for the chirping, is given by
\be
\theta(z,t)\equiv -\frac{s(d_f,t)s^2(z,t)}{2\eta^2 k_c^4\sigma^4(t)}+
\frac{1}{2}\text{arctan}\spare{\frac{s(d_f,t)}{k_c^2\sigma^2_f}}.
\ee

The pulse $E(z,t)$  is shown in \figref{fig:1}(b) at four particular instants of time, taking $k_0>0$ (it propagates rightwards).
At $t=0$ the pulse, centered at $z=0$, is down-chirped, i.e. the wavelength at the front of the pulse is larger than at its tail. As time increases, free evolution in the quadratic dispersion relation induces self-compression of the pulse. Specifically, the width $\sigma(t)$ in \eqnref{sigma} becomes smaller following the decrease of the function $s(d_f,t)$. Maximum compression occurs at a specific time $t_f \equiv \eta d_f/v$, where the width reaches its minimum $\sigma(t_f) = \sigma_f$ and the spatial extension of the pulse is minimized around a compression point $z=d_f$. At this time, all the components of the pulse sync in phase, namely $\theta(z,t_f) = 0$. At latter times $t>t_f$ the pulse becomes up-chirped and it expands in size.
In principle, the compression distance $d_f$ and width $\sigma_f$ of the pulse can be arbitrarily chosen, allowing for deep sub-wavelength compression ($\sigma_f \ll \lambda_0 \equiv 2\pi /k_0$). In \figref{fig:1}(c), we show the mean frequency $\bar\w \equiv \int_\mathds{R}  \w p(\w)\text{d} \w$ and standard deviation $S_\w \equiv [ \int_\mathds{R}  (\w- \bar \w)^2  p(\w)\text{d} \w]^{1/2}$ of the pulse at $z=0$ as a function of the compression width $\sigma_f$, using $p(\w) \equiv |\tilde{E}(0,\w)| / \int_\mathds{R} |\tilde{E}(0,\w)|\text{d}\w  $ with $\tilde E(z,\w) \equiv (2\pi)^{-1/2}\int_\mathds{R}  E(z,t)\exp(-i\omega t) \text{d}t$.
Stronger compression (lower $\sigma_f$) requires higher mean pulse frequencies and wider distributions in frequency space. We consider hereafter sufficiently large spot sizes and small carrier wavenumbers, say $\sigma_f \gtrsim 10^{-1} \lambda_0$ and $k_0 \lesssim 10^{-1}\omega_c/v$, such that the spectral properties of the pulse are consistent with the assumptions considered (\eg~single quadratic band approximation). The frequency distribution of the pulse does not significantly depend on $d_f$.

One can show that $E(z,t)$, as defined above, is consistent within electrodynamics in the medium \eqnref{dispersion}. Indeed, $E(z,t)$ has been constructed as a particular linear combination of electromagnetic field modes, engineered in analogy to the wave-packet contracting quantum dynamics of a massive particle evolving in free space, which  also displays a quadratic dispersion relation (see the supplemental material in \cite{PhysRevLett.109.147205}). The chirped electromagnetic pulses can be produced by driving the waveguide at a given spatial position, say at $z=0$. In \cite{SM} we provide two detailed examples of how the chirped pulses $E(z,t)$ can be engineered in a 3D hollow waveguide with perfectly conducting walls \cite{PhysRevLett.119.043904}, a relevant system for circuit quantum electrodynamics~\cite{PhysRevB.92.174507,Zoepfl}, and in a multi-layer photonic crystal. 

Let us now address the interaction between the self-compressing chirped pulse $E(z,t)$ and a single quantum emitter placed at the position $(x_0,y_0,d)$.  The quantum emitter is first modelled as a qubit with electronic levels $\cpare{ \ket{g},\ket{e}}$ and transition frequency $\omega_{q}$. Its electric dipole moment is assumed to  point along the direction of the component of the electric field considered in $E(z,t)$. 
Accordingly, the Hamiltonian describing the electric-dipole interaction of the qubit with the electromagnetic pulse is given by
\be
\frac{\hat{H}}{\hbar}=\frac{\omega_{q}}{2}\hat{\sigma}_{z}+  \frac{\Omega(d,t)}{2}\hat{\sigma}_{+}+ \frac{\Omega^*(d,t)}{2}\hat{\sigma}_{-}, \label{h-qubit}
\ee
where $\Omega(d,t) \equiv -2d_{eg}E(d,t)/\hbar$ is the time- and position dependent Rabi coupling strength, $d_{eg}$ is the dipole matrix element of the qubit and $\hbar$ the reduced Planck constant. We use the Pauli matrix operators for the qubit levels   $\hat \sigma_z \equiv \ketbra{e}{e}-\ketbra{g}{g}$ and $\hat{\sigma}^+ \equiv [\hat \sigma^-]^\dagger= \ketbra{e}{g}$. 
The dynamics of the state of the qubit $\hat \rho(t)$ including spontaneous emission with rate $\Gamma$ are modeled with the Born-Markov master equation $\partial_t \hat \rho = (\im \hbar)^{-1} [\hat H,\hat \rho] + \Gamma (\hat \sigma_- \hat \rho \hat \sigma_+ - [\hat \sigma_+ \hat \sigma_-, \hat \rho]_+/2)$, which can be numerically solved. 
 We remark that the rotating wave approximation, namely using $\Omega(d,t) \equiv -2 d_{eg}E^+(d,t)/\hbar$ in \eqnref{h-qubit}, can be employed in the regime $\Omega_0 \ll 2\omega_q$ where  $\Omega_0\equiv \max_{d,t }\abs{\Omega(d,t)}$. 
  
\begin{figure}[t]
\noindent \begin{centering}
\includegraphics[scale=1]{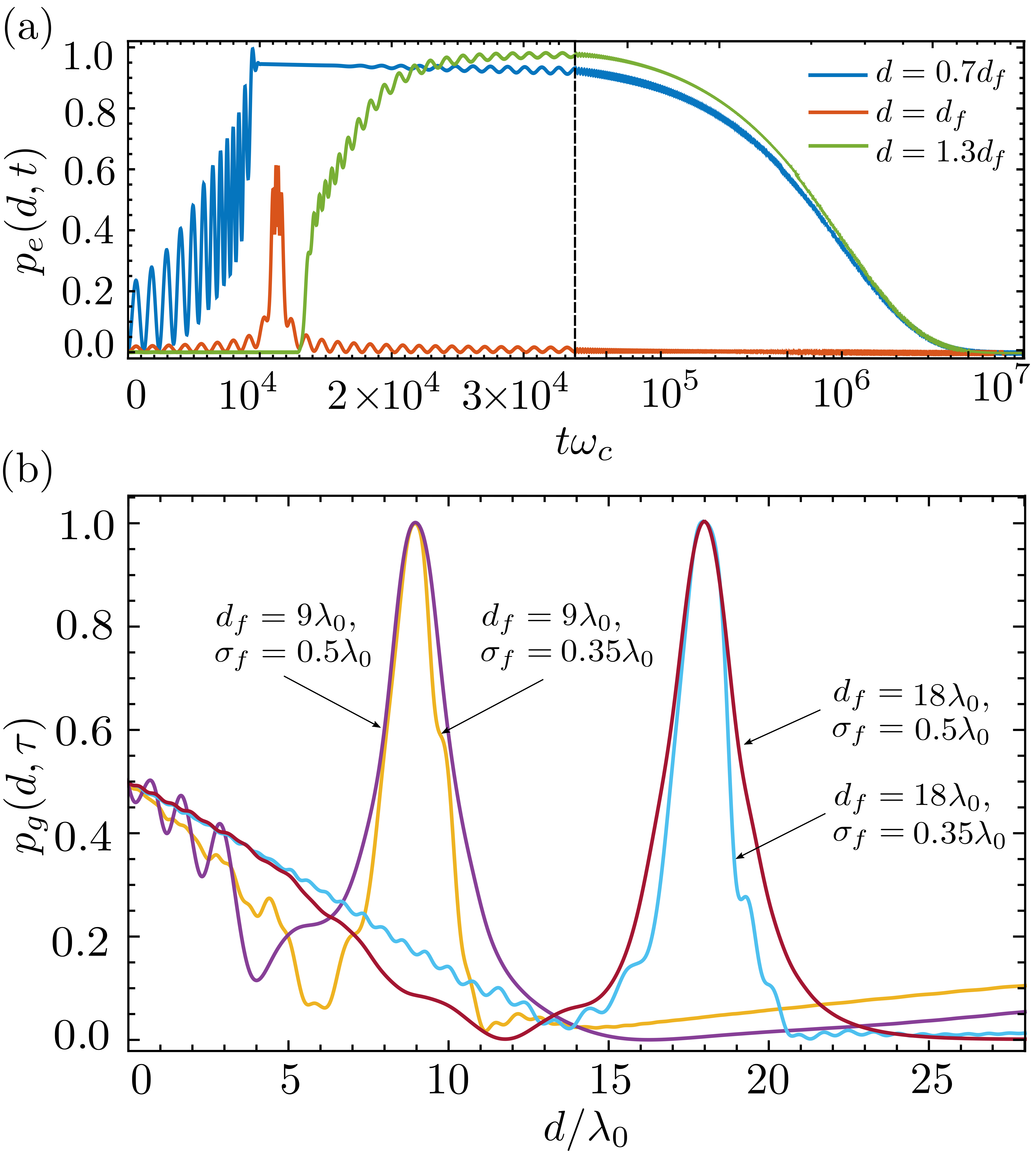}
\par\end{centering}
\caption{ (a) Excited state population of a qubit of frequency $\omega_q=\omega_0$ as a function of time for different positions of the qubit. We choose $\omega_0/\omega_c=1.005$, $\Gamma/\omega_0=10^{-6}$, $d_f/\lambda_0=18$, $\sigma_f/\lambda_0=0.35$, $\Omega_0/\omega_c=0.038$ and $\phi=0$. (b) Ground state population of the qubit as a function of the on-axis distance $d$ from the origin, for different values of $d_f$ and $\sigma_f$ (see inset) and at a time $\tau(d)=2t_f+\eta d/v$ (see main text).  For this panel we fix $\omega_0/\omega_c=1.005$ and $\Omega_0/\omega_c=0.038$ ($\Omega_0/\omega_c=0.030$) for the curves with $\sigma_f/\lambda_0=0.35$  ($\sigma_f/\lambda_0=0.5$).  \label{fig:fig3} }
\end{figure}
Let us analyze the dynamics of a qubit situated at position $z=d,$ and which is initially in the ground state $\hat \rho(0) = \ketbra{g}{g}$.  \figref{fig:fig3}(a) shows the excited state probability $p_e(d,t) = \tr \spare{\hat \rho(t) \ketbra{e}{e}}$ as a function of time for different positions $d$ of the qubit. When the qubit is situated at the compression distance ($d=d_f$), the qubit is excited when the pulse reaches it at $t=t_f$ and de-excited when it travels further away.  Hence, $p_e(d_f,t \gg t_f)\approx 0$. However, when the qubit is situated far from the compression distance ($|d-d_f| \gg \sigma_f$),
it remains excited at long times $p_e(d \neq d_f,t \gg t_f)\approx 1$. The interaction of the qubit with the pulse happens at a timescale shorter than $\Gamma^{-1}$  assuming usual spontaneous emission rates $\Gamma/\omega_q \lesssim 10^{-4}$. \figref{fig:fig3}(b) shows the ground state population of the qubit $p_g(d,t) = \tr \spare{\hat \rho(t) \ketbra{g}{g}}$ as a function of the position $d$ of the qubit, at a time $\tau(d)$ such that $t_f\ll\tau(d)\ll \Gamma^{-1}$, that is, after the interaction with the pulse but before the decay of the qubit. As shown in~\cite{SM}, \figref{fig:fig3}(b) does not depend on $\Gamma$ in the regime $\Gamma/\w_q \lesssim 10^{-5} $.  The plot shows different curves for different values of $d_f$ and $\sigma_f$.  The ground-state population features a peak of height one centered at the compression distance of the pulse $d=d_f$ that is narrower the smaller the value of $\sigma_f$. The peak manifests that the self-compressing chirped pulse prepares a position-dependent state with a spatial resolution $\sigma_q$ (the width of the probability peak) that, as further discussed below, is given by $\sigma_q/\sigma_f \approx 1.34$ and  thus  can be smaller than $\lambda_q \equiv 2 \pi c/ \w_q $. Hence, the proposed self-compressing chirped pulses can be used to perform remote sub-wavelength addressing of quantum emitters with a resolution length scale given by $\sigma_q \propto \sigma_f$.

The dynamics shown in \figref{fig:fig3} can be understood in the context of Landau-Zener (LZ) processes~\cite{LZ,PhysRevA.53.4288}. To this end, we consider the Hamiltonian \eqnref{h-qubit} in the rotating wave approximation and write $\Omega(d,t) \equiv \couplingh \exp \spare{i\vphi}$, where both functions $\couplingh$ and $\vphi$ are real and depend on the amplitude and phase of the electromagnetic pulse, respectively. One then moves to a rotating frame given by the unitary transformation $\hat{U}(t)=\exp \spare{-i\vphi\hat{\sigma}_{z}/2 },$ which converts the Hamiltonian \eqref{h-qubit} into
\be
\frac{\hat{H}_{\text{LZ}}}{\hbar}=\left(\frac{\omega_0}{2}+\detuning\right)\hat{\sigma}_{z}+ \frac{\couplingh}{2} \pare{\hat{\sigma}_{+} +\hat{\sigma}_{-}},\label{eq:H-LZ}
\ee
where $\detuning\equiv \partial_t \vphi /2$ for $\w_q=\w_0$. In \eqcite{eq:H-LZ}, the qubit detuning $\detuning$ (Rabi coupling $\couplingh$) is time-dependent due to the chirping (time-dependent amplitude) of the electromagnetic pulse. The results shown in \figref{fig:fig3} can be explained in the dressed-state picture of \eqnref{eq:H-LZ}. As further illustrated in~\cite{SM}, within the time interval at which the coupling $g>0$ and hence an energy gap opens between the dressed energies, the detuning $\Delta$ undergoes a single change (two changes) of sign whenever the qubit is out of focus $\abs{d-d_f} \gg \sigma_f$ (on focus $\abs{d -d_f} \ll \sigma_f$). 
In 
both regimes the process is adiabatic.  Consequently, the out-of-focus qubit goes forth in the lower dressed-energy branch.
Hence, after the pulse has passed and the energy gap closes ($g=0$),  the qubit ends up in the excited state. However, in the case when the qubit is on-focus, it  goes forth and back in the lower dressed-energy branch, thus ending in the ground state. 
For distances in the crossover regime $|d-d_f| \approx \sigma_f$ the process includes non-adiabatic transitions as the gap closes while the detuning changes sign. By comparing the timescales at which $\Delta$ changes sign and at which the gap opens due to the coupling $g$, we can estimate and numerically verify that the width of the peak in \figref{fig:fig3}(b) is given by the above-quoted value of $\sigma_q /\sigma_f \approx 1.34$ \cite{SM}. 

\begin{figure}[t]
 \noindent \begin{centering}
\includegraphics[scale=0.99]{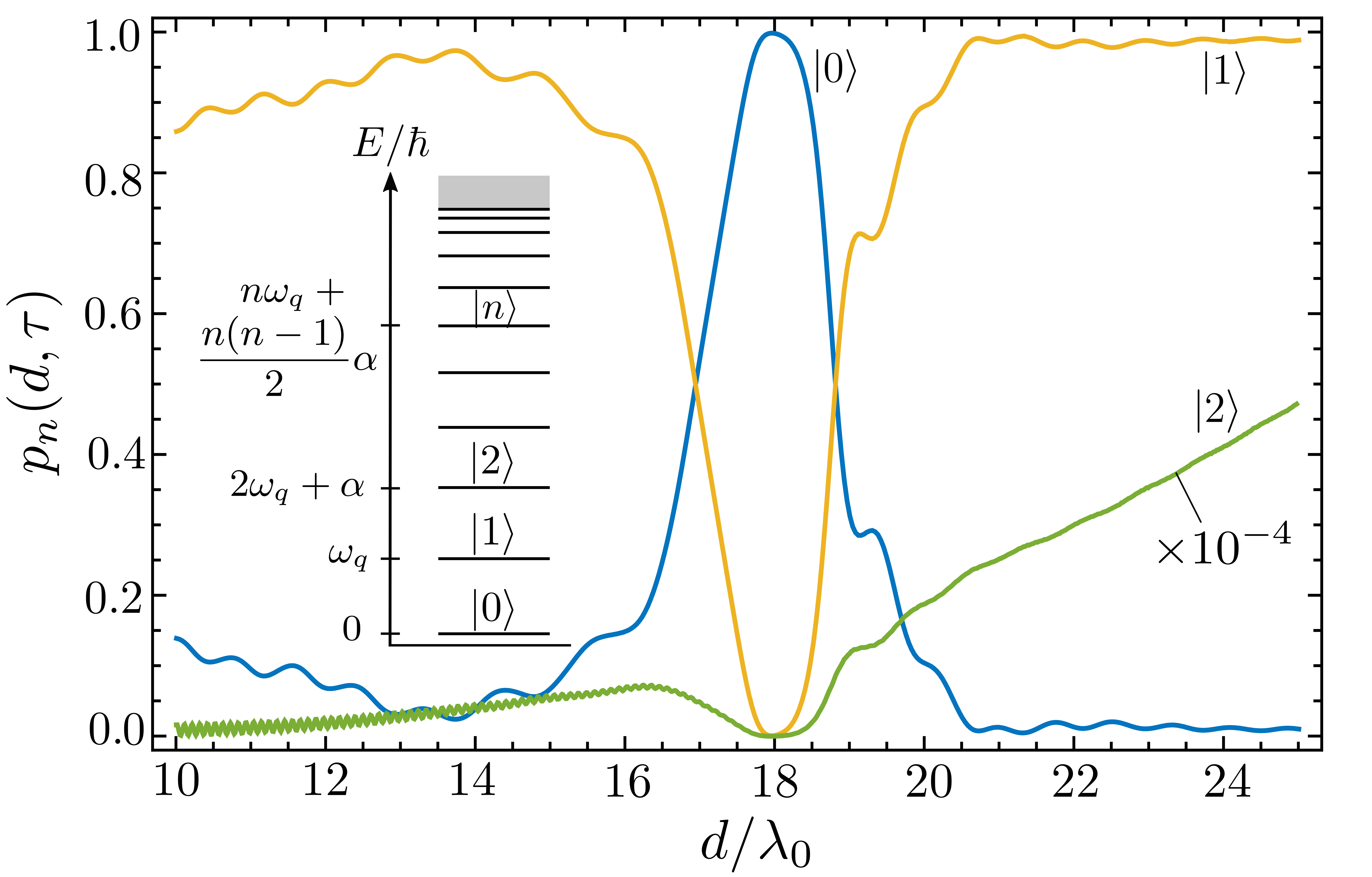}
\par\end{centering}
\caption{ Population of the internal states of an oscillator with  anharmonicity $\alpha/\omega_q=-0.05$ interacting with the pulse as a function of the distance $d$ to the center of the waveguide, at a time $\tau(d)=2t_f+\eta d/v$ such that $t_f\ll\tau(d)\ll \Gamma^{-1}$ with $\Gamma/\omega_q=10^{-6}$.
Parameters used: $\omega_q/\omega_c=\omega_0/\omega_c=1.005$, $d_f/\lambda_0=18$, $\sigma_f/\lambda_0=0.35$, $\Omega_0/\omega_c=0.038$, $\phi=0$. 
\label{fig:anharmonic} }
\end{figure}

Our results also holds for more complex quantum emitters, such as the nonlinear harmonic oscillator that models a transmon qubit \cite{Koch2007}. The Hamiltonian describing the interaction of the quantum emitter with the chirped electromagnetic field pulse is now given by 
\be 
\frac{\hat H}{\hbar} = \omega_\text{q} \hat b^\dagger \hat b + \frac{\alpha}{2}  \left[(\hat b^\dagger \hat b)^2-\hat b^\dagger \hat b\right]+  \frac{\Omega(d,t)}{2} \hat{b}^\dagger+  \frac{\Omega^*(d,t)}{2}\hat{b},
\ee 
where $\hat b$ ($\hat b^\dagger$) is a bosonic annihilation (creation) operator, 
$\alpha$ is the anharmonicity coefficient,   
and $\Omega(d,t) = - 2 d_q E(d,t)/\hbar$, where $d_q$ is the dipole moment of the anharmonic quantum emitter.  One can then numerically solve the Born-Markov master equation  $\partial_t \hat \rho = (\im \hbar)^{-1} [\hat H,\hat \rho] + \Gamma (\hat b \hat \rho \hat b^\dagger - [\hat b^\dagger \hat  b, \hat \rho]_+/2)$ in a truncated sufficiently large Hilbert space. We assume the initial state is $\hat \rho(0) =\ketbra{0}{0}$, where $\hat b^\dagger \hat  b \ket{n} = n \ket{n}$ with $n=0,1,2,\ldots$. In \figref{fig:anharmonic} we plot the excitation probability of the state $\ket{n}$, namely $p_n(t,d) \equiv \bra{n} \hat \rho(t) \ket{n}$, as a function of $d$.  The population of the ground state features a peak around the compression position of the pulse, analogously to the two-level quantum emitter. Note that the asymmetry of the electromagnetic pulse before and after the compression distance is imprinted in the excited states of the anharmonic quantum oscillator.

So far, we have analyzed the interaction of chirped electromagnetic pulses with a single quantum emitter as a function of its position in the waveguide. As shown in \cite{SM}, our results hold
 in the case of an ensemble of many quantum emitters, as illustrated in \figref{fig:1}(a), in the regime where the number of photons in the electromagnetic pulse is much larger than the number of quantum emitters and the single-photon coupling rate is weak. In this regime, both the interactions between quantum emitters and their backaction on
 the electromagnetic pulse, i.e., the total field scattered by the emitters, can be neglected. According to our conservative estimation in \cite{SM}, our results should hold for an ensemble of at least $\approx10$ qubits for the parameters used in Figs. 2 and 3. The regime of few-photon pulses \cite{PhysRevA.86.013811,PhysRevLett.123.123604,alex2020quantum} or strongly coupled quantum emitters, which is notably challenging to approach theoretically due to interesting emerging quantum phenomena \cite{PhysRevLett.120.153602,PhysRevA.92.053834,PhysRevB.79.205111,Mahmoodian_2020} is, in our opinion, a very interesting direction for further research.

Our results are relevant in many platforms where other options such as transverse driving are difficult (e.g. photonic crystals surrounded by a band-gapped medium), harmful (e.g. photodamage in hybrid metal-dielectric waveguides), incapable of individual addressing (e.g. closely packed qubit ensembles), or the source of unwanted decoherence (e.g. in superconducting circuits). The self-compressing behavior described in this work only relies on free propagation in a quadratic dispersion relation, and is thus not specific to the electromagnetic field.
 An interesting outlook of our work is to explore similar self-compressing dynamics in other systems with quadratic spectrum,  collective quasiparticles such as bulk plasmons \cite{QUINN1995460} or exciton-polaritons \cite{nature05131,Ramezani:17}, and even 
  quantum technological platforms such as cavity arrays \cite{lpor.200810046} and atoms in optical lattices \cite{PhysRevX.7.031049}.
By providing new probing and controlling capabilities at the quantum level, self-compressing pulses could thus become a relevant asset for quantum technologies in the future.

We acknowledge discussions with M. L. Juan, G. Kirchmair, A. Sharafiev, and M. Zanner.
C. G. -B. acknowledges funding from the EU Horizon 2020 program under the Marie Skłodowska-Curie grant agreement no. 796725 (PWAQUTEC). J.J.G.-R. acknowledges support form Project No. PGC2018-094792-B-I00 (MCIU/AEI/FEDER, UE), CAM/FEDER Project No. S2018/TCS-4342 (QUITEMAD-CM) and the Quantum Technology Platform PTI-001 (CSIC).

\bibliography{Mybib.bib}

\pagebreak

\newcommand{\beginsupplement}{%
        \setcounter{table}{0}
        \renewcommand{\thetable}{S\arabic{table}}%
        \setcounter{figure}{0}
        \renewcommand{\thefigure}{S\arabic{figure}}%
        \setcounter{equation}{0}
        \renewcommand{\theequation}{S\arabic{equation}}
        \setcounter{page}{1}
     }
\onecolumngrid
\begin{center}
\textbf{\large Supplemental Material}
\end{center}

\beginsupplement
%\makeatletter

\vspace{0.9cm}

\twocolumngrid

\section{Example: hollow 3D waveguide}

Here we describe how to engineer the chirped self-compressing pulse introduced in the main article in a particular waveguide. We consider a hollow cylindrical waveguide of radius $R$ and infinitely extended along $z$ with perfect electric conducting walls \cite{PhysRevLett.119.043904}. The following results are straightforwardly extended to waveguides with other cross-sections. 

\subsection{Electromagnetic field modes}
The electric and magnetic field operators in the waveguide can be expanded in terms of the electromagnetic field modes $\mathbf{f}_\alpha(\rr)$ and eigenfrequencies $\omega_\alpha$, where $\alpha$ denotes a multi-index (to be specified below). In particular, we have
\begin{align}
\hat{\mathbf{E}}(\rr)&=\im \sum_\alpha \sqrt{\frac{\hbar \omega_\alpha}{2\epsilon_0}}\left[\mathbf{f}_\alpha (\rr)\hat{a}_\alpha - \text{H.c.}\right],\label{Efieldmodes}\\
\hat{\mathbf{B}}(\rr)&= \sum_\alpha\sqrt{\frac{\hbar}{2\epsilon_0 \omega_\alpha}}\left[ \nabla\times\mathbf{f}_\alpha (\rr)\hat{a}_\alpha + \text{H.c.}\right]\label{Bfieldmodes},
\end{align}
where $\epsilon_0$ denotes the vacuum permittivity, $\hbar$ denotes the reduced Planck constant, and $\sum_\alpha$ includes the sums (integrals) over discrete (continuous) indices. The creation and annihilation operators, namely $\hat a^\dagger_\alpha$ and  $\hat a_\alpha$ fulfill the commutation relations $[\hat{a}_\alpha,\hat{a}_{\alpha'}^\dagger]=\delta_{\alpha\alpha'}$ and $[\hat{a}_\alpha,\hat{a}_{\alpha'}]=[\hat{a}^\dagger_\alpha,\hat{a}_{\alpha'}^\dagger]=0$. Here $\delta_{\alpha\alpha'}$ contains a Kronecker (Dirac) delta for each discrete (continuous) index.

The electromagnetic field modes and eigenfrequencies can be determined by solving the eigenmode equation
\be
\nabla\times\nabla\times \mathbf{f}_\alpha(\rr) - \epsilon(\mathbf{r}) \frac{ \w_\alpha^2}{c^2}\mathbf{f}_\alpha(\rr)=0 \label{vector-wave},
\ee
where $\epsilon(\mathbf{r})=1$, and the boundary condition for perfect electric conducting walls  $\mathbf{f}_\alpha(\rr)\times \mathbf{e}_r = 0$  at $|\rr|=R$. Here $c$ denotes the speed of light in vacuum and $\mathbf{e}_r$ denotes the radial unit vector. It can be shown \cite{modesWaveguide} that the electromagnetic field modes split up into two families, namely the transverse electric $(s=\text{TE})$ and the transverse magnetic $(s=\text{TM})$ modes. The electric (magnetic) field of the TE (TM) modes is transverse to the direction of propagation. Furthermore, the modes are denoted by two discrete indices $n\in\mathds{N}_0$ and $m\in\mathds{N}$ that characterize the azimuthal and radial distribution of each mode respectively. The continuous index $k\in\mathds{R}$ denotes the projection of the wave-vector on the symmetry axis of the waveguide and completes the multi-index $\alpha\equiv( s,n,m,k)$ which uniquely characterises each mode $\mathbf{f}_\alpha(\rr)=\mathbf{f}_{nm}^s(k;\mathbf{r})$.

The dispersion relation for the eigenfrequencies is given by
\be
\omega_\alpha=\omega_{nm}^s(k)\equiv c\sqrt{(k_{nm}^s)^2+k^2},\label{dispersion-all}
\ee
where $k_{nm}^\text{TM} \equiv p_{nm}/R$ and  $k_{nm}^\text{TE} \equiv q_{nm}/R$. The constants $p_{nm}$ and $q_{nm}$ denote the $m$-th root of the $n$-th order Bessel function of the first kind $J_n(x)$ and the $m$-th root of $\partial_x J_n(x)$ respectively. Note that each tuple $(s,n,m)$, denoted as $s_{nm}$, characterises an energy band in $k$. Let us now focus on the lowest TM band, namely $\text{TM}_{01}$. The modes in this band are characterized by a a non-zero electric field component on axis. For modes with $k\lesssim \omega_c/(2v),$ the dispersion relation can be approximated by the quadratic dispersion relation
\be
\dispersion \equiv \omega_{01}^\text{TM}(k)\approx \omega_c+\frac{v^2}{2\omega_c}k^2.\label{SP:dispersion}
\ee 
Here $\omega_c \equiv c p_{01}/R$ and $v \equiv c$. In this regime, the contribution to the dispersion relation of the non-quadratic terms is less than 1\%. Note that higher-order (higher frequency) bands can always be neglected for electromagnetic modes whose frequencies are below the lower cutoff of any such band, namely $cp_{nm}/R$. Specifically, since the lowest energy band above the $\text{TM}_{01}$ band is the $\text{TM}_{11}$ band, one can neglect any contribution from higher order bands for modes with frequency $\omega < c p_{11}/R$. Note that the above inequality can always be fulfilled for a sufficiently small radius $R$. As shown later, the quadratic approximation is satisfied if the pulse parameters satisfy the condition $\left(k_{0}+2/\sigma_{f}\right)\lesssim\omega_c/(2v)$.

Let us turn our attention to the corresponding electric field operator. In particular, the expectation value of the z-component of this operator evaluated on the axis of the waveguide is given by
\be 
\langle\hat{E}_z(z,t)\rangle=i\int_\mathds{R}\sqrt{\frac{\hbar\dispersion}{2\epsilon_{0}}}\left[\wmodesz\langle\ak(t)\rangle-\hc\right]\text{d}k,\label{electricfield}
\ee
where the evolution of the operators $\hat{a}(k)\equiv \hat{a}_{01}^\text{TM}(k)$ are given by $\hat{a}(k;t)= \hat{a}(k) \exp[-\im \omega(k) t]$. Moreover, 
\be
\wmodesz \equiv \mathbf{f}^\text{TM}_{01}(k;z)\cdot \mathbf{e}_z=\frac{i C}{\omega(k)}e^{ikz},\label{eq:modes}
\ee
denotes the $z$-component of the corresponding field mode along the symmetry axis. Here $C$ is a real constant that depends on the geometry of the waveguide. In particular, for a cylindrical geometry,  
\be
C=\frac{\omega_{c}}{R\sqrt{2\pi^{2}J_{1}^{2}\left(\omega_{c}R/c\right)}}.
\ee

\subsection{Preparation of the chirped pulse} 
Let us assume that the modes of the waveguide are prepared in a coherent state with an amplitude $\amplitude$ given by
\be
\amplitude=C_\alpha \sqrt{\omega(k)} e^{i\phi}e^{-\frac{1}{2}\sigma_f^2(k-k_0)^2}e^{i \frac{d_f}{2k_0}(k-k_0)^2}. \label{amplitude}
\ee
The amplitude $\alpha(k)$ is centered around $k_0$ and has a width given by $1/\sigma_f$. Hence, the modes width a wavenumber $k\in[k_0-2/\sigma_f,k_0+2/\sigma_f]$ contribute to the electric field. Here $C_\alpha$ is a dimensional constant that can be related to the number of photons in the waveguide, $N_\text{ph}$, through 
\be
N_\text{ph}=C^2_{\alpha} \sqrt{\pi } \frac{c^2 \left(2 k_0^2 \sigma_f^2+1\right)+4 \sigma_f^2 \omega_c^2}{4 \sigma_f^3 \omega_c}.
\ee
Evaluating \eqcite{electricfield} with this coherent state leads to the definition of the chirped pulse $E(z,t)$ introduced in the main article. The pulse amplitude $N$ defined in the main text can then be related to $C_\alpha$ through  $N=C_\alpha C k_c \sqrt{\pi \hbar/\epsilon_0}$. 
%Notice that the amplitude depends on the momentum $k$ of the mode. 
The particular form of the amplitude $\alpha(k)$ has been chosen in analogy to the 
wave-packet contracting quantum dynamics of a massive particle evolving in free space, which also follows a quadratic dispersion relation. In particular, it is equivalent to the momentum representation of the wavepacket that describes a massive particle with an additional initial imaginary phase that causes it to contract (see supplemental material in \cite{PhysRevLett.109.147205}).

In order to prepare the considered chirped pulse, we must prepare each mode of the waveguide with an amplitude given by the multimode coherent state in \eqcite{amplitude}. 
To this end, we apply a time-dependent driving at a point $z_0$ along the axis of the waveguide, that we set to $z_0=0$.  The Hamiltonian of the driven waveguide  is in that case 
\be
\frac{\hat{H}_\text{D}}{\hbar}=\int_\mathds{R}  \omega(k) \hat{a}^\dagger(k)\hat{a}(k)\text{d}k+\driving\hat{a}(z_{0})+\drivingc\hat{a}^{\dagger}(z_{0}), \label{eq:h-driving1}
\ee
Here $\driving$ is a complex function that describes the time-dependent driving. The operators $\hat{a}^\dagger(z_{0})$ and $\hat{a}(z_{0})$  create or annihilate a photon at the position $z_{0}$, and are related to the operators $\hat a(k)$ 
through
\be
\hat{a}({z_0})=\frac{1}{\sqrt{2\pi}}\int_\mathds{R} \ak e^{i k z_0}\text{d}k .
\ee
To obtain the required driving function $\driving$, we calculate the equations of motion for the expected values $\langle\hat{a}(k)\rangle$ using the Hamiltonian \eqcite{eq:h-driving1}, and impose that they are equal to the amplitudes in \eqcite{amplitude} after the driving pulse, i.e. at times such that $\driving = 0$. Under the assumption that the driving function $D(t)$ is extended over a sufficiently long time interval, we obtain a condition for the driving in spectral representation, i.e., for the Fourier transform of the driving function $D(t)$, $\tilde{D}(\omega)\equiv(2\pi)^{-1}\int_\mathds{R} D(t)\exp(-i\omega t)\text{d}t$. Specifically, the driving $\tilde{D}(\omega)$ has to fulfill the condition 
\be
\tilde{D}(\dispersion)=i\alpha^*(k). \label{driving-alpha}
\ee
Since $\omega(k)$ is a continuous function with an image in $[\omega_c,\infty)$, this condition fixes the value of the driving for the relevant frequencies $\omega\geq\omega_c$.

Since the Fourier transform of the driving $\driving$ is proportional to the momentum distribution of the engineered pulse (see Eq.~\eqref{driving-alpha}), it displays similar properties as the ones discussed in Figure 1(c) in the main text, namely it contains components of increasingly high frequency for increasingly compressed pulses (i.e. for smaller values of $\sigma_f$). Such high frequency components increase the experimental demands required to engineer the driving pulse, ultimately limiting the compression capabilities. To estimate these limitations, we calculate the electric field profile obtained by removing its high-frequency components above an upper cutoff $\omega_r$, i.e., by setting $\tilde{E}(z,\omega > \omega_r) = 0$, where $\tilde E(z,\w) \equiv (2\pi)^{-1/2}\int_\mathds{R}  E(z,t)\exp(-i\omega t) \text{d}t$. The cutoff frequency $\omega_r$ is determined by the accessible frequencies in the laboratory. The electric field pulse generated by such frequency-truncated spectral distribution is given by
\be
E_\text{truncated}(z,t)=\frac{1}{\pi}\int_\mathds{R} E(z,s) \frac{\sin\left[\omega_r(t-s)\right]}{t-s}\text{d}s,
\ee
where $E(z,s)$ is the original electromagnetic field pulse (Eq. 2 in the main text). 
For the range of $\sigma_f$ and $d_f$ considered in the main article ($d_f/\lambda_0\gtrsim 10$ and $\sigma_f/\lambda_0\lesssim0.5\lambda_0$), one can show that
this truncated field remains a good approximation for 
$E(z,t)$ for upper cutoffs as low as $\omega_r=2\omega_c$ or, in other words, the electric field profile is not significantly modified after removing its higher frequency components. This suggests that the self-compressing pulses could be realistically engineered in a hollow waveguide, provided that the applied driving approximates well enough the condition in \eqcite{driving-alpha}. 

\subsection{Classical field approximation and extension to multiple qubits}

Throughout the article we have considered the electric field as a classical variable, \ie, we have assumed $\hat{a}(k)=\langle \hat{a}(k)\rangle+\delta \hat{a}(k)\approx \langle\hat{a}(k)\rangle$. The contribution of the quantum fluctuations to the interaction with a single qubit can be estimated by the quantity $\Omega_q\equiv[\langle\hat{\Omega}_{\text{q}}^{\dagger}(d,t)\hat{\Omega}_{\text{q}}(d,t)\rangle]^{1/2}$, where the operator is $\hat{\Omega}_{\text{q}}(d,t)$ is defined as
\be
\hat{\Omega}_{\text{q}}(d,t)=-id_{eg}\int_{\mathbb{R}}\sqrt{\frac{\hbar \omega(k)}{2\epsilon_0}} f_z(k;z)e^{-i\omega(k)t}\hat{a}(k)\text{d}k,
\ee
and the expected value is calculated on a thermal state with temperature $T$. For the high intensity pulses considered here, the value of $\Omega_q$ is several orders of magnitude smaller than the classical coupling at an arbitrary time and position inside the waveguide. Therefore, one can neglect the quantum fluctuations in the interaction with the pulse with a single qubit.

We now consider an ensemble of $N_\text{q}$ qubits of internal frequencies $\omega_\text{q}$ situated in an array along the symmetry axis of the waveguide. The dynamics of the emitters are independent if the following conditions are fulfilled: first, if the classical field approximation is valid for each single qubit, and secondly, if the electric field scattered by each qubit can be neglected. We have already proved that the first condition holds. Let us now look at the latter. We denote by $U_0$ the energy of the incoming pulse. The energy scattered by each single qubit, denoted by $U_\text{sc}$, is upper bounded by $U_\text{dipole}$, where $U_\text{dipole}$ is the energy radiated by a classical dipole oscillating at the transition frequency $\omega_\text{q}$ situated at the compression distance of the pulse. The power $P$ radiated by such a dipole inside a hollow cylindrical waveguide can be calculated using the formalism of dyadic Green functions \cite{chen_to_tai}, which yields
\be
P=\frac{\left|d_{eg}\right|^{2}c}{4\pi\epsilon_{0}R^{4}}\frac{p_{01}^{2}}{J_{1}^{2}(p_{01})}\frac{\omega_{\text{q}}}{\sqrt{\omega_{\text{q}}^{2}-\omega_{c}^{2}}},
\ee
where $d_{eg}$ is the amplitude of the dipole moment. The energy radiated by the oscillating dipole is then $U_\text{dipole}=P\times\Delta \tau,$ where $\Delta \tau$ is the interaction time with the pulse, estimated by the full width at half maximum (FWHM) of the coupling function at $d=d_f$,
\be
\Delta \tau=\frac{4\eta\sigma_{f}\sqrt{2\log\left(2\right)}}{c\sqrt{1-\frac{2\eta^{2}}{k_{c}^{2}\sigma_{f}^{2}}\log\left(2\right)}},
\ee
where $k_c=\omega_c/c$ and $\eta$, $\sigma_f$ are pulse parameters. For the pulse used in Figs. 2 and 3 of the main text, $\nu\equiv U_\text{dipole}/U_0\simeq0.079.$ Therefore, the fraction of energy scattered by each qubit to the pulse is $U_\text{sc}/U_0<\nu\ll 1$. The dynamics of $N_\text{q}+1$ emitters are then decoupled provided that the total energy scattered by the previous $N_\text{q}$ qubits remains much smaller than the energy of the incoming pulse, namely $N_\text{q}\nu\ll1$. As a consequence, we predict that our model applies for ensembles of at least $N_\text{q}\approx10$ qubits. We remark that this is a conservative estimation based on the upper bound for the scattered energy.

%\pagebreak

\section{Example: multi-layer photonic crystal}

\begin{figure}
\begin{centering}
\includegraphics[scale=1]{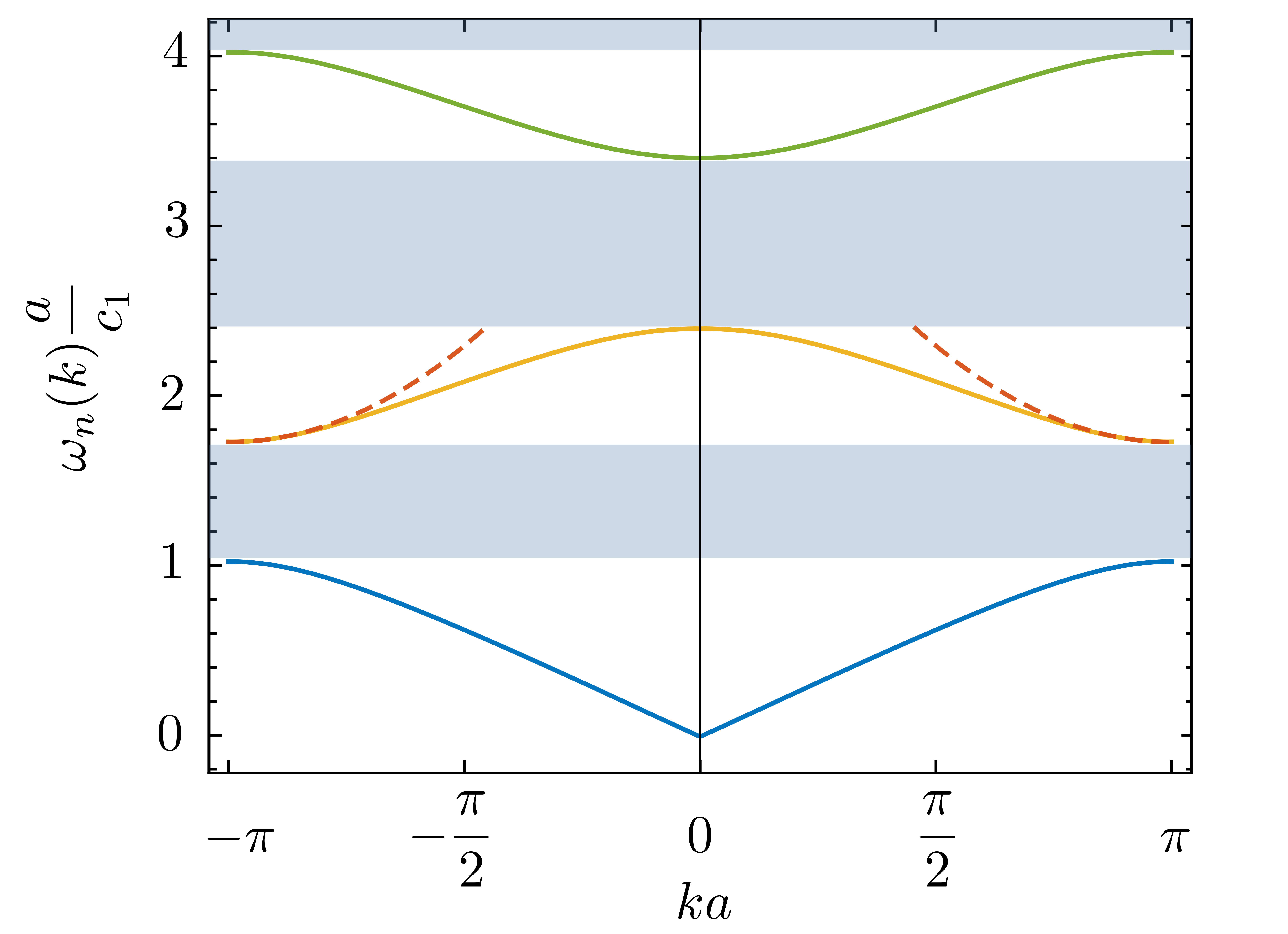}
\par\end{centering}
\caption{Dispersion relation $\omega_{n}(k)$ of a multi-layer film in the first Brillouin zone, showing the first three bands (solid lines). Quadratic
approximation of the dispersion relation of the second band (dashed
line). Parameters of the photonic crystal: $c_{2}=0.3c_{1},$ $\epsilon_{1}=11.1\epsilon_{2},$
$b=0.5a$, $v\simeq0.88c_1$.\label{fig:dispersion-layer}}
\end{figure}

In this section we analyse how to implement the chirped self-compressing pulse in a particular photonic crystal.  The implementation 
follows analogous steps as the one introduced in the previous section. We consider an infinite multi-layer photonic crystal constituted by a material with linear response to electromagnetic fields, characterized by a relative permeability $\mu=1$ and a relative permittivity $\epsilon(\mathbf{r})$, defined by
\be
\epsilon(\mathbf{r})=\begin{cases}
\epsilon_{1} & \text{for }0\leq z<b\\
\epsilon_{2} & \text{for }b\leq z<a,
\end{cases}\label{permittivity-crystal}
\ee
where the pattern is repeated along the $z$ direction, \ie $\epsilon(\mathbf{r}+na\mathbf{e}_z)=\epsilon(\mathbf{r})$, where $n\in \mathbb{Z}$. The length $a$ denotes the unit cell of the crystal, and $b$ satisfies $b<a$. We note that the results can be extended to photonic crystals with other periodic translation symmetries.

The electric and magnetic field operator inside the photonic crystal can be expressed in terms of the electromagnetic field modes in the same way as introduced in \eqcite{Efieldmodes} and \eqcite{Bfieldmodes}. In this case, the electromagnetic field modes $\mathbf{f}_\alpha (\mathbf{r})$ are the solutions to the eigenmode equation in \eqcite{vector-wave}, with the permittivity given by \eqcite{permittivity-crystal} and the boundary conditions $\mathbf{e}_z\times\left(\mathbf{f}_{\alpha}^2-\mathbf{f}_{\alpha}^1\right)=0$ at each layer interface. Due to the continuous translation invariance of the system along the $x$ and $y$ directions, and the discrete translation invariance along the $z$ direction, the eigenmodes are Bloch functions \cite{Joannopoulos}, \ie, they are of the form
\be
\mathbf{f}_\alpha (\mathbf{r})=e^{ik z}\mathbf{u}_\alpha(z),
\ee
where we have set $\mathbf{k}=k\mathbf{e}_z$ for modes propagating along the $z$ direction. Here $\mathbf{u}_\alpha(z)$ is a periodic function of the lattice period $a$,  $\mathbf{u}_\alpha(z+la)=\mathbf{u}_\alpha(z)$, for $l\in \mathbb{Z}$. 
For the multi-layer crystal, the eigenmodes are characterized by the multi-index $\alpha=(s,n,k)$, and can be denoted by $\mathbf{f}_\alpha(\mathbf{r})=\mathbf{f}_{s,n}(k;\mathbf{r})$, where $k\in \mathbb{R}$ is the wavenumber of the mode along the $z$ direction,  $n\in \mathbb{N}$ is the band index, and $s=\pm$ denotes two degenerate modes for each pair $(n,k)$ \cite{layered}. The dispersion relation of the modes is given by the implicit equation \cite{Joannopoulos}
\be
\tan\left(ka\right)=\sqrt{\frac{4}{\alpha^{2}(\omega)}-1,}\label{eq:dispersion-multilayer}
\ee
where
\begin{align}
\alpha(\omega)&=2\cos\left(\frac{b\omega}{c_{1}}\right)\cos\left(\frac{\omega(a-b)}{c_{2}}\right)\\
&-\frac{\left(c_{1}^{2}+\text{\ensuremath{c_{2}}}^{2}\right)\sin\left(\frac{b\omega}{c_{1}}\right)\sin\left(\frac{\omega(a-b)}{c_{2}}\right)}{c_{1}\text{\ensuremath{c_{2}}}},
\end{align}
and the velocities $c_{1},$ $c_{2}$ represent the group velocity
within each layer, defined as $c_{1}\equiv c/\sqrt{\epsilon_{1}}$, $c_{2}\equiv c/\sqrt{\epsilon_{2}}$. \figref{fig:dispersion-layer} shows the band structure in the first Brillouin zone stemming from \eqcite{eq:dispersion-multilayer}. One can see that the dispersion relation of the second band is quadratic around $k\simeq\pi/a$, and can thus be approximated by
\be
\omega_{2}(k)=\omega_{c}+\frac{1}{2}\frac{v^{2}}{\omega_{c}}\left(k-\frac{\pi}{a}\right)^{2},
\ee
 where $\omega_c$ is the cutoff frequency of the second band and $v$ is the curvature. More specifically, the quadratic approximation holds for $k\in[\pi/a-3/4,\pi/a+3/4]$, \ie, in this range the quadratic dispersion differs by less than 1\% from the exact one.

\begin{figure}
\noindent \begin{centering}
\includegraphics[scale=1]{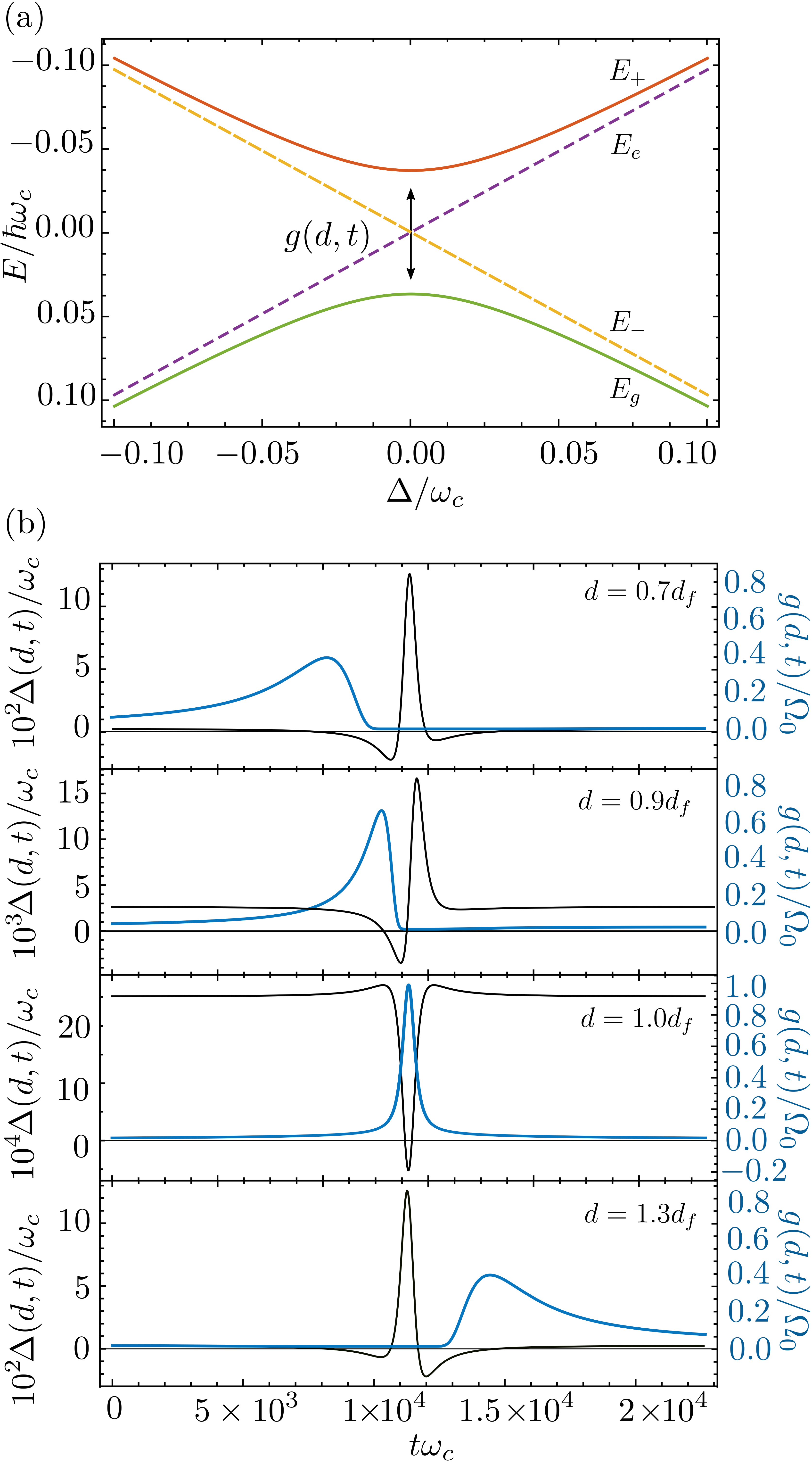}
\par\end{centering}
\caption{(a) Energies of the bare $(E_g,E_e)$ and dressed states $(E_\pm)$ of the qubit in the Landau-Zener process in \eqcite{eq:H-LZ-sup}. (b) Detuning (left axis) and coupling strength (right axis) as a function of time for 4 different positions of the qubit (see inset). Parameters used:  $\omega_0/\omega_c=1.005$, $d_f/\lambda_0=18,$ $\sigma_f/\lambda_0=0.35$, $\Omega_0/\omega_c=0.038$. \label{SP:LZ}}
\end{figure} 

The electric field inside the photonic crystal restricting to the dynamics within the second band then reads
\be
\hat{\mathbf{E}}(\mathbf{r},t){=}
i\sum_{s}\!\!\int_{-\pi/a}^{\pi/a}\!\sqrt{\frac{\hbar\omega_{2}(k)}{2\epsilon_0}}\!\left[\mathbf{f}_{s,2}(k;\mathbf{r})\hat{a}_{s,2}(k)(t)\!
-\!\hc\right]\!\text{d}k\label{Efield-photonic-crystal},
\ee
where $\mathbf{f}_{s,2}(k;\mathbf{r})=e^{ikz}\mathbf{u}_{s,2}(k;z)$.  In a similar fashion to the previous section, we assume that the modes with $s=+$ of the photonic crystal are prepared in a coherent state with an amplitude $\alpha(k)$ given by
\be
\alpha\left(k\right)=\frac{1}{\sqrt{\omega_{2}(k)}}e^{i\phi}e^{-\frac{\sigma_{f}^{2}}{2}\left(k-(k_{0}+\pi/a)\right)^{2}}e^{i\frac{d_{f}}{2k_{0}}\left(k-(k_{0}+\pi/a)\right)^{2}},\label{alpha_photonic}
\ee
where $k_0$, $d_f$ and $\sigma_f$ are free parameters, while the modes with $s=-$ are not populated. The expected value of the electric field is then given by \eqcite{Efield-photonic-crystal} with $\langle\hat{a}_{s,2}(k)(t)\rangle=\alpha(k)\exp{(-i\omega_2(k)t)}$. Notice that then the expression of $\langle\hat{\mathbf{E}}(\mathbf{r},t)\rangle$ has a similar form to that in \eqcite{electricfield}, with the important difference that the modes have an additional position dependence.

If the function $\mathbf{u}_{+,2}(k;z)$ varies slowly compared to the amplitude $\alpha\left(k\right)$,
it is possible to apply the envelope approximation, mainly to approximate
$\mathbf{u}_{+,2}(k;z)\simeq\mathbf{u}_{+,2}(k_0+\pi/a;z)$, where $k_0+\pi/a$ is the carrier wavenumber of the pulse in \eqcite{alpha_photonic}. Under this approximation, the evolution of the field corresponds to a chirped pulse that self-compresses at a position $d_f$ with a spot size $\sigma_f$, multiplied by a position dependent function. The maximum compression happens at a time $t_f=\eta d_f/v$, where $\eta\equiv k_c/(k_0+\pi/a)$ and $k_c\equiv\omega_c/v$. One can show numerically that the envelope approximation holds in the parameter regime where the pulse lies in the quadratic part of the dispersion relation of the second band. For the parameters of the photonic crystal shown in \figref{fig:dispersion-layer}, this corresponds to pulses with $k_0 a \lesssim 10^{-1}$  and $\sigma_f/a \gtrsim 3$. The minimum spot size compared to the free space wavelength of the emitter,  $\lambda_\text{q}=2\pi v/\omega_c$, is then lower bounded by $\sigma_f/\lambda_\text{q} \gtrsim 1$.

\section{Additional Figures}

\begin{figure}
\noindent \begin{centering}
\includegraphics[scale=0.99]{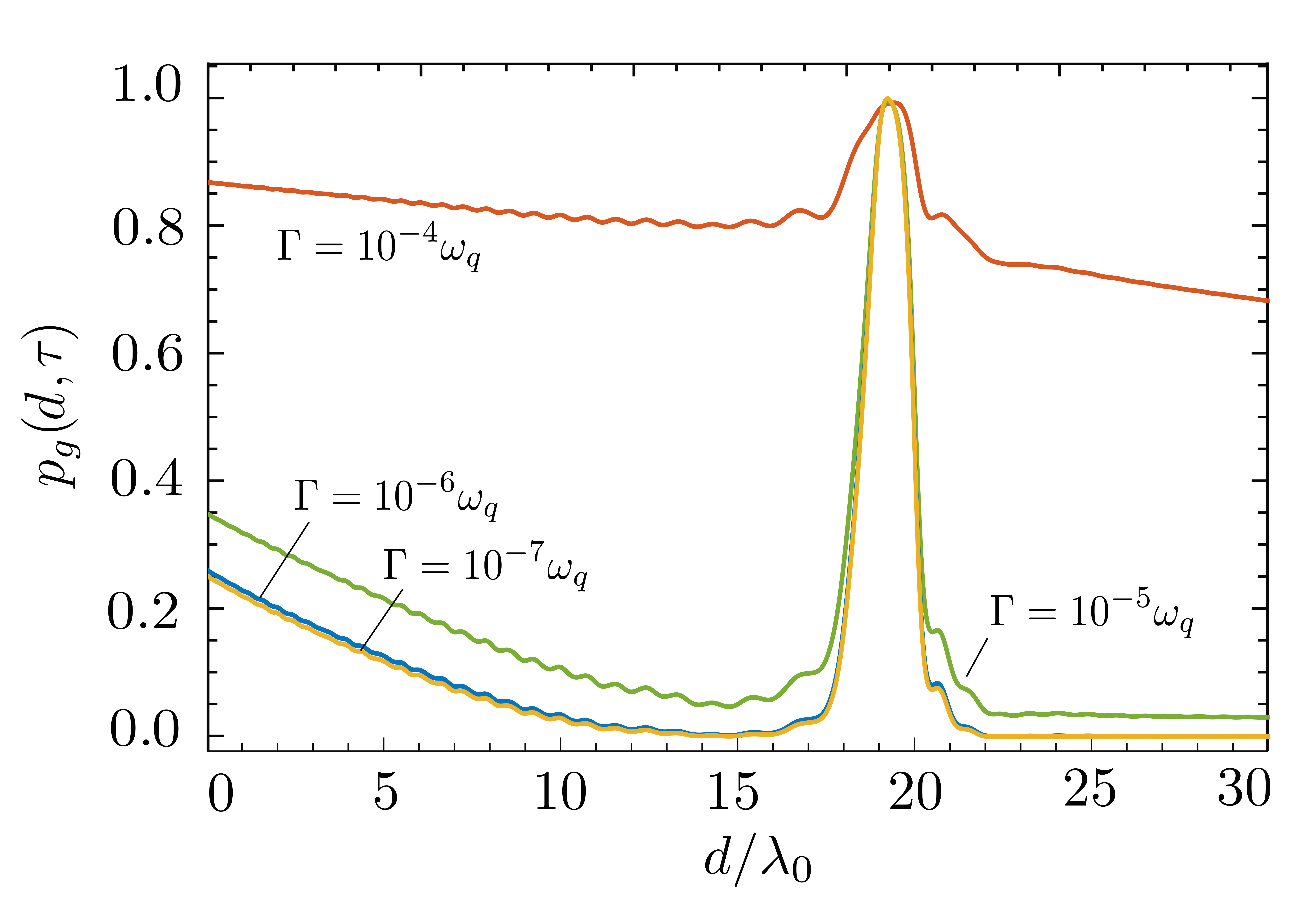}
\par\end{centering}
\caption{Ground state population of the qubit after the interaction with the pulse ($\tau(d)=2t_f+\eta d/v$) as a function of the distance $d$ to the center of the waveguide, for different qubit decay rates $\Gamma$.  Parameters used: 
 $\omega_q/\omega_c=\omega_0/\omega_c=1.005$, $d_f/\lambda_0=18,$ $\sigma_f/\lambda_0=0.35$, $\Omega_0/\lambda_0=0.038$. \label{fig:dissipation} }
\end{figure}

 \figref{SP:LZ}(a) shows the instantaneous eigenenergies of the Landau-Zener Hamiltonian 
 \be
\frac{\hat{H}_{\text{LZ}}}{\hbar}=\left(\frac{\omega_0}{2}+\detuning\right)\hat{\sigma}_{z}+ \frac{\couplingh}{2} \pare{\hat{\sigma}_{+} +\hat{\sigma}_{-}},\label{eq:H-LZ-sup}
\ee
 as a function of the detuning $\Delta(d,t)$. The energies feature an avoided crossing around $\Delta(t,d)=0$ with a gap proportional to the coupling strength $g(t,d)$. However, since the coupling strength is time-dependent and only has positive values for a certain time interval, the gap is not always open. When the gap is closed ($g(t,d)=0$), the eigenstates coincide with the two internal states of the qubit, and their energies are equal to the bare energies (dashed lines in the figure). 
 \figref{SP:LZ}(b) shows the time dependence of the detuning $\Delta(d,t)$ and the coupling strength $g(d,t)$ due to the interaction between the quantum emitter and the self-compressing pulse, for different positions $(x_0,y_0,d)$ of the emitter. In the figures one can see that the relation between the two relevant timescales, namely the timescale at which the detuning changes sign and the timescale at which the gap opens and closes, depend critically on the qubit position, leading to the dynamics explained in the main article.
The opening and closing times of the gap, denoted by $t_o$ and $t_c$, can be estimated by the extremes of the FWHM of the function $g(d,t)$. The  width $\sigma_\text{q}$ is defined as $\sigma_\text{q}\equiv d_2-d_1$, where $d_1$ ($d_2$) is the position for which $t_c$ ($t_o$) coincides with the change of sign of the detuning $\Delta(d,t)$. One can numerically show that the width $\sigma_\text{q}$ is directly proportional to the spot size of the pulse, $\sigma_\text{q}/\sigma_f\approx 1.34$.

Figure \ref{fig:dissipation} shows the ground state population of a qubit placed at the position $(x_0,y_0,d)$ after the interaction with the pulse, for different qubit spontaneous decay rates $\Gamma$ in the range $\Gamma/\omega_q\in[10^{-7},10^{-4}]$. One can see that even for decay rates as large as $\Gamma/\omega_q=10^{-4}$, there is a significant imprint of the self-compressing pulse on the population of a qubit placed at the compression point $z=d_f$. The robustness of such imprint against qubit loss stems from the fast qubit-pulse interaction, which happens at much shorter timescales than the dissipation of the qubit.

%\pagebreak
\end{document}